\documentstyle[12pt]{article}
\newcommand{\ap}{\mbox{$\alpha ^{\prime }$}}
\begin{document}
\begin{flushright}
SJSU/TP-93-10\\January 1993\end{flushright}
\vspace{1.7in}
\begin{center}\Large{\bf On the Definition of Decoherence}\\
\vspace{1cm}
\normalsize\ J. Finkelstein\footnote[1]{Participating Guest, Lawrence
        Berkeley Laboratory \\
       \hspace*{\parindent} \hspace*{1em}
        e-mail: JFINKEL@SJSUVM1.BITNET}\\
        Department of Physics\\
        San Jos\'{e} State University\\San Jos\'{e}, CA 95192, U.S.A
\end{center}
\begin{abstract}
We examine the relationship between the decoherence of quantum-mechanical
histories of a closed system (as discussed by Gell-Mann and Hartle)
 and environmentally-induced diagonalization
of the density operator for an open system.  We study a definition of
decoherence which incorporates both of these ideas, and show that it leads
to a consistent probabilistic interpretation of the reduced density operator.
\end{abstract}
\newpage
\section{Introduction}
In classical physics, there is no apparent need to analyze the role
of the observer.  As long as one assumes that observations can
 be made with negligible effect on the properties observed, one can
ascribe properties to objects without considering whether or not those
properties have actually been observed.
In quantum physics, at least in the Copenhagen interpretation,
the situation is radically different: objects are assumed not to
 have any properties at all, unless and until those properties
have been measured by an outside observer.  Under such an assumption,
an analysis of measurement becomes an indispensable part of
the interpretation of the theory.

In quantum cosmology the ``object'' of study is the entire universe.
The quantum theory of the universe must be interpreted without
relying on the idea of measurement.  This is not, as was the case in
classical physics, because measurement is innocuous; rather it is because
it  is impossible.  There can not be any outside observer of
the entire universe!

It has been suggested \cite{gh1,gh2} by Gell-Mann and Hartle (denoted
by GH in the following) that the Copenhagen notion of measurement
could be replaced by the concept of  ``decohering histories''.
This concept is a generalization of the idea of ``consistent
histories'' advanced by Griffiths \cite{gr} and studied also by
Omn\`{e}s \cite{om}; a recent paper by Dowker and Halliwell \cite{dh}
analyzes several example of decohering histories.  GH suggest several
different versions of the condition for decoherence (which they
refer to as the weak, medium, medium-strong, and strong conditions),
and it is an open question within the program of GH of what is the
best definition of decoherence to impose.

Other authors \cite{z} have investigated how, when a system interacts
with its environment, the density operator for the system becomes
diagonal (in a particular basis).  This diagonalization can also be
argued to provide an answer to the question of how, without relying
on the idea of measurement, it is possible to say when a property is
real or when an event has happened; in fact this diagonalization
has also been termed decoherence.  In the following, we shall refer
to the diagonalization of the density operator for a system, due to
its interaction with its environment, as Z decoherence.

In this paper we study the properties of yet another definition of
the decoherence of histories, which we will refer to as PT decoherence.
We will see that PT decoherence implies the medium decoherence
condition of GH, and that it also involves the diagonalization of the
density operator; in a sense PT decoherence can be taken to
characterize those histories which decohere by the mechanism
of Z decoherence.  In the next section we will review the definitions of
histories and decoherence as they were used in refs \cite{gh1,gh2} and
\cite{dh}; we will then define a generalization of the decoherence
functional, and use it to state the condition that
we call PT decoherence.  In the final section we will establish and
discuss some of the properties of histories which exhibit PT decoherence,
and compare them with properties implied by other definitions.
\section{Formalism}
Since we are interested in those sets of histories for which we can
discuss Z decoherence, we will limit our discussion to the case in which,
 from among all the variables which describe the world, we distinguish
a certain fixed subset of them, and say that this subset describes
the ``system''; the remaining variables then describe the ``environment''.
Formally we write the state-space $\cal H$ of the world as the tensor
product of spaces $\cal S$ of the system and $\cal E$ of the
environment: ${\cal H}={\cal S}\otimes {\cal E}$.  The discussion of
histories will involve (Schr\"{o}dinger picture) projection operators $P$,
and we only consider those projection operators which act trivially
on $\cal E$, and so can be written
\mbox{ $P=P_{\cal S}\otimes I_{\cal E}$.}

Following GH, we define a history by a sequence of projections
made at a fixed sequence of times $t_{1}\cdots t_{n}$ which satisfy
$t_{1}<\cdots <t_{n}$. For each given time $t_{k}$, the set of projection
operators will be denoted by $\{ P^{k}_{\alpha _{k}}\}$ , where a
particular value of the index $\alpha _{k}$ denotes a particular operator
in the set. The superscript $k$ on $P$ is required since we are
 allowing different sets of projectors at
different times.  For each fixed value of $k$, the projections are supposed
to represent an exhaustive set of exclusive alternatives, which implies
 that these operators satisfy
 $P_{\alpha }P_{\beta }=\delta _{\alpha \beta}P_{\alpha}$ and
 $\sum_{\alpha} P_{\alpha }=I$.
A particular history is then labelled by a sequence
 $\alpha = [\alpha_{1},\cdots ,\alpha _{n}]$.  For a particular sequence
$\alpha $, define
\begin{equation}
C_{\alpha }=P^{n}_{\alpha _{n}}e^{-iH(t_{n}-t_{n-1})}P^{n-1}_{\alpha
_{n-1}}e^{-iH(t_{n-1}-t_{n-2})}\cdots P^{1}_{\alpha _{1}}
e^{-iH(t_{1}-t_{0})} .
\end{equation}
The condition of the world at the initial time $t_{0}<t_{1}$
 will be represented by the density
operator $\rho (t_{0})$.  Then the decoherence
functional is \cite{gh1,gh2}
\begin{equation}
D(\ap ,\alpha)={\rm Tr}[C_{\alpha
^{ \prime}}\rho (t_{0})C_{\alpha }^{\dagger }] .
\end{equation}
The definition of decoherence that was emphasized in \cite{gh2} was the
``medium decoherence'' condition; this is the requirement that, for two
alternative histories labelled by $\alpha $ and $\ap $,
\begin{equation}
D(\ap ,\alpha)=0 .
\end{equation}
We will say that a family of histories
for which Eq. (3) holds satisfies ``GH decoherence''.
Probabilities $p(\alpha )$ can be assigned to the members of a family of
histories satisfying Eq. (3) by $p(\alpha )=D(\alpha ,\alpha )$;
Eq. (3) then guarantees that $p(\alpha\: {\rm or}\: \beta )=p(\alpha )
+p(\beta )$, where $p(\alpha\: {\rm or}\: \beta)$ is calculated from
Eq. (2) with $C_{\alpha\: {\rm or}\: \beta }=C(\alpha )+C(\beta )$.

One would certainly expect that, in cases where the GH decoherence
condition (Eq. (3)) is (at least approximately) satisfied, this would
come about because tracing over the environment in Eq. (2) would give
(at least approximately) zero; this is well illustrated by examples
considered in \cite{gh2} and in \cite{dh}.  In order to formalize this
expectation, we now define a generalized decoherence functional $\bar{D}$
by restricting the trace in Eq. (2) to be just over the environment;
that is, we define
 \begin{equation}
\bar{D}(\ap ,\alpha )={\rm Tr}_{\cal E}[C_{\alpha^{ \prime }}
\rho (t_{0})C_{\alpha }^{\scriptscriptstyle \dagger }] .
 \end{equation}
For fixed $\alpha $ and $\ap $, $\bar{D}(\ap ,\alpha )$
is thus an operator in $\cal S$, which can easily be shown to satisfy
\begin{eqnarray}
{\rm Tr}\bar{D}(\ap ,\alpha ) & = & D(\ap ,\alpha ) \\
\bar{D}(\alpha ,\ap) & = & \bar{D}^{\dagger }(\ap  ,\alpha ) \\
\sum _{\alpha^{\prime} , \alpha }\bar{D}(\alpha \prime ,\alpha ) & = &
\rho _{\cal S}(t_{n})
 \end{eqnarray}
where in Eq. (7) $\rho _{\cal S}$ is the density operator for the system:
$\rho _{\cal S}={\rm Tr}_{\cal E}[\rho ]$, and we have used
$\sum _{\alpha }C_{\alpha}
=e^{-iH(t_{n}-t_{0})}$.  We can now state the condition that
we call (since it is defined by a Partial Trace) PT decoherence:
we will say a family of histories satisfies PT decoherence if,
for any alternative histories $\alpha$ and $\ap $,
\begin{equation}
\bar{D}(\ap ,\alpha )=0 .
 \end{equation}

In the next section we will establish and discuss the following properties
of PT decoherence:
\begin{enumerate}
\item PT decoherence implies the medium-decoherence condition of GH.
\item For a family of histories satisfying PT decoherence, the projectors
at the final time $t_{n}$ must be, in the space $\cal S$, onto
(perhaps a coarse graining of) a basis which diagonalizes
 $\rho _{\cal S}(t_{n})$.
In this sense a PT-decoherent family of histories satisfies Z decoherence.
\item For each history $\alpha $ in a decoherent family of histories,
one can define an ``effective'' density operator for the system
$\rho_{\cal S}^{\alpha }$; PT decoherence then implies, and is essentially
implied by, a consistency condition for these effective density
operators: $p(\alpha \: {\rm or}\: \beta )\rho_{\cal S}^{\alpha \:
{\rm or}\: \beta }=p(\alpha )\rho ^{\alpha }_{\cal S}+p(\beta )
\rho ^{\beta }_{\cal S}$.
\item If $\rho$, the density operator for the world, represents a pure state,
then GH decoherence implies \cite{gh1,gh2} the existence of
``generalized records'' of the histories.  PT decoherence then implies
further that these generalized records exist {\em in the environment}.
\item To the extent to which it is possible to neglect the interaction
between the system and the environment for times later than $t_{n}$,
histories extended past $t_{n}$ satisfy the following property:
any two histories which are PT-decoherent at $t_{n}$ will continue to
be PT-decoherent at all later times, for any choice of projections at
these later times.  In the case where $\rho $ represents a pure state,
this would imply the {\em persistence} of records. \end{enumerate}
\section{Implications of the Formalism}
\hspace{\parindent} \underline{First}, any two histories
 which are PT-decoherent (Eq. (8)) are necessarily
GH-decoherent (Eq. (3)). This follows from Eq. (5).

\underline{Second}, to discuss a relationship
 between PT decoherence and Z decoherence,
suppose we sum a family of histories over all projectors at all times earlier
than $t_{n}$; if $\alpha$ represents such a summed history, it follows
 from Eq.~(1) that $C_{\alpha}=P^{n}_{\alpha _{n}}e^{-iH(t_{n}-t_{0})}$.
Since all of our projectors act trivially in $\cal E$, we can write
$P^{n}_{\alpha _{n}}=P_{{\cal S}\alpha }\otimes I_{\cal E}$, where
$P_{{\cal S}\alpha }$ is a projection operator acting on $\cal S$.
So if $\alpha $ and $\ap$ are two such histories, we have
\begin{eqnarray}
\bar{D}(\ap ,\alpha ) & = & {\rm Tr}_{\cal E}[(P_{{\cal S}\alpha^{\prime}}
\otimes
I_{\cal E})e^{-iH(t_{n}-t_{0})}\rho (t_{0})e^{iH(t_{n}-t_{0})}
(P_{{\cal S}\alpha }\otimes I_{\cal E})] \nonumber \\*
 & = & P_{{\cal S}\alpha^{\prime}}
{\rm Tr}_{\cal E}[\rho (t_{n})]P_{{\cal S}\alpha}
\nonumber \\*
 & = & P_{{\cal S}\alpha^{\prime} }\rho _{\cal S}(t_{n})P_{{\cal S}\alpha } .
\end{eqnarray}
Thus, since when
$\ap \neq \alpha$, PT decoherence requires $\bar{D}(\ap ,\alpha)=0$ ,
 Eq. (9) implies in this case
\begin{equation}
P_{{\cal S}\alpha^{\prime} }\rho _{\cal S}(t_{n})P_{{\cal S}\alpha }=0 .
\end{equation}
So $\rho_{\cal S}(t_{n})$ has no matrix elements between the subspaces
of $\cal S$ projected onto by $P_{{\cal S}\alpha^{\prime} }$
 and $P_{{\cal S}\alpha }$.
If each member of the set $\{ P_{{\cal S}\alpha }\} $ projects onto a
one-dimensional subspace of $\cal S$, i.e. if $P_{{\cal S}\alpha}=
|\alpha\rangle \langle \alpha |$,
 then Eq. (10) obviously says that $\rho _{\cal S}
(t_{n})$ is diagonal in the basis $\{ |\alpha\rangle \}$.
  More generally, Eq. (10)
says that $\rho _{\cal S}(t_{n})$ is block-diagonal in the subspaces
projected onto by the $\{ P_{{\cal S}\alpha }\} $; since it is always possible
to diagonalize $\rho _{\cal S}(t_{n})$ on each such subspace separately,
this means that there exists a basis in which $\rho _{\cal S}(t_{n})$
is diagonal and such that each $P_{{\cal S}\alpha }$ is a sum of projectors
onto that basis.

This establishes that the set of projectors for the final time $t_{n}$
of a PT-decoherent family necessarily project onto (possibly a
coarse-graining of) a basis which diagonalizes $\rho _{\cal S}(t_{n})$.
Therefore, at time $t_{n}$ a PT-decoherent family also exhibits Z
decoherence.  However, at earlier times the situation is not so simple.
If we consider, for example, two histories $\alpha $ and $\ap $ for which
the only non-trivial projections are at $t=t_{n-1}$, we could write,
in place of Eq. (9),
\begin{equation}
\bar{D}(\ap ,\alpha)={\rm Tr}_{\cal E}[e^{-iH(t_{n}-t_{n-1})}
(P_{{\cal S}\alpha^{\prime }}\otimes I_{\cal E})\rho (t_{n-1})
 (P_{{\cal S}\alpha }\otimes
I_{\cal E})e^{iH(t_{n}-t_{n-1})}] .
\end{equation}
The exponential factors in Eq. (11) do not necessarily cancel because
${\rm Tr}_{\cal E}$ is not cyclic in operators which act upon $\cal S$.
Hence we can \underline{not} conclude that, in analogy with Eq. (10),
$P_{{\cal S}\alpha^{\prime }}\rho_{\cal S}(t_{n-1})P_{{\cal S}\alpha}=0$,
which would be the Z-decoherence condition at $t=t_{n-1}$.

\underline{Third}, for each history $\alpha$ in a GH-decoherent family one can
define an ``effective'' density operator $\rho^{\alpha}$ by
\begin{equation}
\rho^{\alpha}=\frac{C_{\alpha}\rho(t_{0})C_{\alpha}^{\dagger}}{{\rm
 Tr}[C_{\alpha}\rho(t_{0})C_{\alpha}^{\dagger}]} .
\end{equation}
Using this effective
density operator corresponds, in the case of a
pure state, to the ``collapse of the state-vector''. We can also define
the effective density operator for the system by
\begin{eqnarray}
\rho_{\cal S}^{\alpha} & \equiv & {\rm Tr}_{\cal E}[\rho^{\alpha}]
\nonumber \\*
 & = & \frac{{\rm Tr}_{\cal E}[C_{\alpha}\rho(t_{0})C_{\alpha}^{\dagger}]}{{\rm
Tr}[C_{\alpha}\rho(t_{0})C_{\alpha}^{\dagger}]}
\end{eqnarray}
which implies
\begin{equation}
\bar{D}(\alpha,\alpha)=p(\alpha)\rho_{\cal S}^{\alpha} .
\end{equation}
Now let $\alpha$ and $\beta$ represent two alternative histories; from
Eq. (4) with $C_{\alpha \: {\rm or}\: \beta}=C_{\alpha}+C_{\beta}$ we get
\begin{equation}
\bar{D}(\alpha \: {\rm or}\: \beta,\alpha \: {\rm or}\: \beta)=\bar{D}
(\alpha,\alpha)+\bar{D}(\beta,\beta)+\bar{D}(\alpha,\beta)+\bar{D}
(\beta,\alpha) .
\end{equation}
If $\alpha$ and $\beta$ represent alternative
members of a PT-decoherent family,
the last two terms in Eq. (15) vanish; then Eqs. (14) and (15) imply
\begin{equation}
\rho^{\alpha \: {\rm or}\: \beta}_{\cal S}=
\frac{p(\alpha)}{p(\alpha)+p(\beta)}\rho_{\cal S}^{\alpha}
+\frac{p(\beta)}{p(\alpha)+p(\beta)}\rho_{\cal S}^{\beta} ,
\end{equation}
or if we sum over {\em all} alternative members of the family, we get
\begin{equation}
\rho_{\cal S}=\sum _{\alpha}p(\alpha)\rho_{\cal S}^{\alpha} .
\end{equation}
Eq. (16) is the consistency condition for the effective density operator
which is implied by PT decoherence.  A necessary and sufficient condition
for Eq.~(16) is that the sum of the last two terms in Eq. (15) vanishes,
for any alternative histories $\alpha$ and $\beta$; by using Eq. (6),
we can write this condition as
\begin{equation}
\bar{D}(\alpha,\beta)+\bar{D}^{\dagger}(\alpha,\beta)=0 .
\end{equation}
This condition is somewhat weaker than is PT decoherence; it implies
the weak, but not the medium, decoherence condition defined by GH.

\underline{Fourth}, if the density operator $\rho$ represents a pure state,
 we can write
$\rho(t_{0})=|\Psi\rangle \langle \Psi |$. The decoherence functional
defined in Eq. (2) is then $D(\ap ,\alpha)=\langle\Psi|C_{\alpha}^{\dagger
}C_{\alpha^{\prime}}|\Psi\rangle$, and the GH decoherence condition
is that, for $\alpha$ and $\ap $ representing alternative histories,
$C_{\alpha}|\Psi\rangle$ and $C_{\alpha^{\prime}}|\Psi\rangle$ are
orthogonal.  The fact that alternative histories lead, at time
$t=t_{n}$,
to orthogonal states implies what GH refer to as ``generalized records''
of the histories. The PT-decoherence condition becomes, in the case of
a pure state,
\begin{equation}
{\rm Tr}_{\cal E}[C_{\alpha^{\prime}}|\Psi\rangle\langle \Psi|C_{\alpha}
^{\dagger}]=0 .
\end{equation}
It can be shown that
Eq. (19) is equivalent to the condition that there exist subspaces
${\cal E}_{\alpha}$ and ${\cal E}_{\alpha^{\prime}}$ of $\cal E$ such that
$C_{\alpha}|\Psi\rangle\in S\otimes {\cal E}_{\alpha}$,
$C_{\alpha^{\prime}}|\Psi\rangle\in S\otimes {\cal E}_{\alpha^{\prime}}$,
and ${\cal E}_{\alpha}$
 is orthogonal to ${\cal E}_{\alpha^{\prime}}$. We can call this
condition ``orthogonality in the environment''
\footnote{If ${\cal E}_{\alpha}$ and/or ${\cal E}_{\alpha^{\prime}}$
has dimension greater than one, it is possible to make a finer graining
(perhaps in a branch-dependent way).  For our discussion it is simpler
not to assume this has been done.}.

In the case in which $\rho $ represents a pure state, the
PT-decoherence condition is
therefore equivalent to the statement that
$C_{\alpha}|\Psi\rangle$ and $C_{\alpha^{\prime}}|\Psi\rangle$
are orthogonal in the environment.  We can thus say that a
PT-decoherent family of histories produces records in the environment.

\underline{Fifth}, the histories we have discussed so far extend until a
time $t_{n}$, which we can call the ``present''. Suppose we now consider
further extending these histories to a time $t_{m}>t_{n}$, the ``future''.
Let $\alpha$ represent, as before, a sequence of alternatives at times
$t_{1}\cdots t_{n}$, and let $\beta$ represent a sequence at times
$t_{n+1}\cdots t_{m}$.  Then if we denote by $(\alpha$ and $\beta)$
the sequence consisting of first $\alpha$ and then $\beta$, we can write,
using an obvious extention of the notation introduced in Eq.\ (1),
$C_{\alpha\: {\rm and}\: \beta}=C_{\beta}C_{\alpha}$. The generalized
decoherence functional becomes
\begin{equation}
\bar{D}(\ap \: {\rm and}\: \beta^{\prime}
,\alpha\: {\rm and}\: \beta)={\rm Tr}_{\cal E}
[C_{\beta^{\prime}}C_{\alpha^{\prime}}\rho(t_{0})C_{\alpha}^{\dagger}
C_{\beta}^{\dagger}].
\end{equation}
Now let us suppose that we could ignore the interaction between the
system and the environment for all times in the future. Then, for future
times, we could write the Hamiltonian $H$ of the world as a sum:
$H=H_{\cal S}\otimes I_{\cal E}+I_{\cal S}\otimes H_{\cal E}$. This means
that $C_{\beta}$ could be written
\begin{equation}
C_{\beta}=C_{{\cal S}\beta}\otimes e^{-iH_{\cal E}(t_{m}-t_{n})} ,
\end{equation}
where $C_{{\cal S}\beta}$ is an operator on $\cal S$; there would be
a similar expression for $C_{\beta^{\prime }}$. Then Eq. (20) would imply
\begin{eqnarray}
\bar{D}(\ap \: {\rm and}\: \beta^{\prime},\alpha\: {\rm and}\: \beta) &
= & C_{{\cal S}\beta^{\prime}}{\rm Tr}_{\cal E}[C_{\alpha^{\prime}}
\rho(t_{0})C_{\alpha}^{\dagger }]C^{\dagger }_{{\cal S}\beta} \nonumber \\
&=&C_{{\cal S}\beta^{\prime}}\bar{D}(\ap ,\alpha)C^{\dagger }_{{\cal S}\beta} .
\end{eqnarray}
If the original histories $\alpha$ and $\ap$ are PT-decoherent, then
$\bar{D}(\ap ,\alpha)$ vanishes and so the extended histories are
PT-decoherent also. Then, (under the assumption that we could ignore
interaction between the system and the environment in the future), we
see that two histories which are PT-decoherent in the present will continue
to be PT-decoherent when extended into the future; this is true for any
choice of projection in the future.  In the case in which $\rho$ represents
 a pure state, GH decoherence, and so {\em a fortiori} PT
decoherence, implies the existence of records; then if PT decoherence
persists, the records will persist also.

Of course it is completely unrealistic to expect that a system would
not interact with its environment in the future.  Rather, we would expect
that interaction would continue, but that coherence, once lost to the
environment, would never be recovered by the system.  However, this
last expectation is not guaranteed by the formalism we are using
 (although the authors cited in \cite{z} might consider that it should be
part of the definition of Z decoherence); we have not specified, for
example, that the environment be large.

\underline{Finally}, let us summarize the relationship between the
three kinds of decoherence we have discussed, in the special case in which
$\rho$ represents a pure state.  In this case, GH decoherence implies
the existence of ``generalized records'', i.e.\ orthogonal states at the
end-points of each history.  However, these records might not be in the
environment, where they would be expected to persist.  Perhaps a family
of histories which satisfied GH decoherence, but not PT decoherence,
should be called a family of ``orthogonal histories''. Z decoherence,
on the other hand, does describe a mechanism by which records of the
present state of the system will appear in the environment.  However,
histories are more general than is the present state of a system;
for example two clearly distinguishable histories, with records set in
stone out there in the environment, might not be Z-decoherent if they
happened to lead to the same state of the system.  PT decoherence
distinguishes those sets of histories for which there are records
in the environment.

Consider again the two states $C_{\alpha}|\Psi \rangle $ and
$C_{\alpha^{\prime }}|\Psi \rangle $. If they are orthogonal, they are
GH-decoherent; if they are orthogonal in the environment, they are
PT-decoherent; if they are orthogonal both in the system and in the
environment, they are Z-decoherent.

\vspace{1cm}
\begin{flushleft}
\large {\bf Acknowledgements}
\end{flushleft} \normalsize
\vspace{0.5cm}

I have benefited from conversations with Henry Stapp about the program
of Gell-Mann and Hartle.  I appreciate also the hospitality of the
Lawrence Berkeley Laboratory, where this work was done.

\end{document}